\documentclass[amsmath, amssymb, nofootinbib, nobibnotes, aps, prd]{revtex4-2}

\usepackage{mathtools}
\usepackage{bm}
\usepackage{siunitx}
\usepackage{slashed}
\usepackage{physics2}
\usephysicsmodule{braket}
\usepackage{comment}
\usepackage{graphicx}
\usepackage{xcolor}
\usepackage{orcidlink}
\usepackage{placeins}
\usepackage{float}
\usepackage{hyperref}

\DeclareMathOperator{\Tr}{Tr}

\begin{document}
\title{Hidden-Charm Tetraquarks in a Mixture Model: Coupled-Channel Analysis with $c\bar{c}$ and Hadronic Molecular Components}
\author{Kotaro Miyake\,\orcidlink{0009-0004-1563-9453}}%
\email{miyake@hken.phys.nagoya-u.ac.jp}%
\affiliation{Department of Physics, Nagoya University, Nagoya 464-8602, Japan}
\author{Yasuhiro Yamaguchi\,\orcidlink{0000-0003-1347-0821}}%
\email{yamaguchi@hken.phys.nagoya-u.ac.jp}%
\affiliation{Department of Physics, Nagoya University, Nagoya 464-8602, Japan}%
\affiliation{Kobayashi-Maskawa Institute for the Origin of Particles and the Universe, Nagoya University, Nagoya, 464-8602, Japan}

\begin{abstract}
The nature of the $X(3872)$ and other exotic hadrons has been a subject of extensive investigation since the first observation of the $X(3872)$ in 2003. While various theoretical models have been proposed, including hadronic molecular and compact tetraquark interpretations, some experimental evidence suggests that the $X(3872)$ may be a mixture state of a hadronic molecule and a $c\bar{c}$ core.
In this work, we perform a systematic study of the hidden-charm tetraquark candidates $X(3860)$, $X(3872)$, and $Z(3930)$ using a coupled-channel model that incorporates both $c\bar{c}$ states and $D^{(*)}\bar{D}^{(*)}$ hadronic molecular components. The $c\bar{c}$ sector is described based on the constituent quark model predictions for the $\chi_{cJ}(2P)$ ($J = 0, 1, 2$) states, while the meson-meson interactions are modeled using pseudoscalar and vector meson exchange potentials.
The model parameters are fixed to reproduce the masses of the $X(3872)$ and $Z(3930)$, and the resulting framework is used to predict the mass and structure of the $J^{PC} = 0^{++}$ state associated with the $X(3860)$. Our results support the mixture interpretation of these exotic hadrons, exhibiting strong attractions from the transition potential between $c\bar{c}$ and $D^{(*)}\bar{D}^{(*)}$ components. The molecular component is found to dominate in the $X(3872)$, while the $c\bar{c}$ component plays a more prominent role in the $X(3860)$ and $Z(3930)$.
\end{abstract}

\maketitle

\section{Introduction}\label{sec:intro}

Exotic hadrons have garnered significant interest in the fields of nuclear and hadron physics, as they cannot be described within the conventional quark model picture, i.e., baryons as three-quark states and mesons as quark-antiquark pairs. The possibility of such states was already anticipated in the early stages of quark model development~\cite{Gell-Mann:1964ewy,Zweig:1964ruk, Zweig:1964jf}. In recent years, numerous experimental studies have reported candidates for exotic hadrons, particularly those containing charm quarks, such as $XYZ$, $T_{cc}$, and $P_c$~\cite{Brambilla:2019esw,Yamaguchi:2019vea,Chen:2022asf,ParticleDataGroup:2024cfk}.

Despite the wealth of new data, the internal structures of these exotic hadrons remain poorly understood. One promising framework is the multiquark picture, where compact configurations such as tetraquarks (four-quark states) and pentaquarks (five-quark states) are studied using few-body calculation techniques. Alternatively, near hadron thresholds, molecular interpretations---where exotic hadrons are regarded as loosely bound states of conventional hadrons~\cite{Voloshin:1976ap,DeRujula:1976zlg,Tornqvist:1993ng}---are often favored over compact multiquark configurations.
Indeed, many observed exotic states are located near relevant thresholds: for example, some $XYZ$ states near the $D\bar{D}^{*}$ threshold, $T_{cc}$ near the $DD^*$ threshold, and the $P_c$ states near the $\bar{D}^{(*)}\Sigma_c$ thresholds. However, the underlying hadron-hadron interactions that are responsible for the formation of such molecular states are still not well understood.
Understanding the structure of exotic hadrons and the nature of hadron-hadron interactions is essential for exploring the nonperturbative dynamics of low-energy Quantum Chromodynamics (QCD).

The $X(3872)$\footnote{$X$ and $Z$ states introduced in this paper are shown as $\chi_{cJ}$ in the table of Particle Data Group, where $J$ is the total angular momentum of the states.} is one of the most prominent exotic hadrons in the charm sector, first reported by the Belle experiment in 2003~\cite{Belle:2003nnu}. Since its discovery, the $X(3872)$ has been confirmed by various experimental groups~\cite{CDF:2003cab,D0:2004zmu,BaBar:2004oro,LHCb:2011zzp,CMS:2013fpt,BESIII:2013fnz,ATLAS:2016kwu}, establishing it as a well-studied candidate for an exotic hadronic state. It was observed near the $D^0\bar{D}^{*0}$ threshold, which has raised significant interest in its internal structure.
The mass of the $X(3872)$ cannot be easily explained within the conventional quark model interpretation as a $c\bar{c}$ charmonium state. Instead, it is widely considered to be a tetraquark state composed of $c\bar{c}q\bar{q}$, where $q$ denotes a light quark $(q=u,d)$. The quantum numbers of the $X(3872)$ have been determined as $J^{PC} = 1^{++}$~\cite{LHCb:2013kgk,LHCb:2015jfc}, further constraining its possible structure.

An intriguing feature of the $X(3872)$ is its decay pattern: the decay ratios of $X(3872) \to \omega J/\psi$ and $X(3872) \to \pi^+\pi^- J/\psi$ are found to be comparable, which suggest a significant strong isospin violation~\cite{BaBar:2010wfc,BESIII:2019qvy,LHCb:2022jez}. 
This observation provides additional evidence that the $X(3872)$ does not fit into the conventional charmonium spectrum and supports interpretations beyond the traditional quark model framework.

The $X(3872)$ is located very close to the $D^0\bar{D}^{*0}$ threshold, which has motivated extensive studies of its structure as a hadronic molecule~\cite{Voloshin:2003nt,Close:2003sg,Wong:2003xk,Swanson:2003tb,Tornqvist:2004qy,Swanson:2004pp,Voloshin:2004mh,Voloshin:2003nt,Close:2003sg,Wong:2003xk,Swanson:2003tb,Tornqvist:2004qy,Swanson:2004pp,Voloshin:2004mh,Suzuki:2005ha}. However, experimental observations suggest that the $X(3872)$ may not be a purely molecular state. In particular, evidence from prompt production in high-energy collisions~\cite{Esposito:2015fsa,Esposito:2016noz,Olsen:2017bmm} and radiative decays such as the ratio of the branching fraction of $X(3872) \to J/\psi\gamma$ to that of  $X(3872) \to \psi(2S)\gamma$~\cite{BaBar:2008flx,Belle:2011wdj,LHCb:2014jvf,BESIII:2020nbj} indicates the possible presence of a compact $c\bar{c}$ component.

One possible $c\bar{c}$ component of the $X(3872)$ is the $\chi_{c1}(2P)$ state, which has been predicted by constituent quark models~\cite{Godfrey:1985xj,Barnes:2005pb}, while has not yet been observed experimentally. Its mass is predicted to lie several tens of MeV above the $D^0\bar{D}^{*0}$ threshold. Based on this, theoretical studies have investigated the $X(3872)$ as a mixture state composed of the $\chi_{c1}(2P)$ core and a $D\bar{D}^*$ molecular component~\cite{Takizawa:2012hy,Matheus:2009vq,Danilkin:2010cc,Coito:2010if,Coito:2012vf,Takizawa:2012hy,Ortega:2012rs,Ferretti:2013faa,Chen:2013pya,Takeuchi:2014rsa,Yamaguchi:2019vea,Kinugawa:2023fbf,Terashima:2023tun}.

In addition to the hadronic molecule and mixture state approaches, numerous studies have investigated the structure of the $X(3872)$ using alternative frameworks such as compact tetraquark models~\cite{Maiani:2004vq,Ebert:2005nc,Terasaki:2007uv,Dubnicka:2010kz,Kim:2016tys} and Lattice QCD calculations~\cite{Yang:2012mya,Prelovsek:2013cra,Padmanath:2015era}. 
Heavy exotic hadrons, including the $X$ states, have been extensively studied using a variety of theoretical approaches. For further details, see also the references cited in Refs.~\cite{Brambilla:2019esw,Yamaguchi:2019vea,Chen:2022asf}).

Several experimental candidates for spin partners of the $X(3872)$ have also been reported. 
For the $J^{PC} = 0^{++}$ channel, the $X(3860)$ and $X(3915)$ have been observed near the $D^{(*)}\bar{D}^*$ thresholds. The $X(3860)$ has been reported in the $e^+e^- \to J/\psi D\bar{D}$ process by the Belle collaboration~\cite{Belle:2017egg}, while the obtained mass and widths have a large ambiguity, and the state is not seen in the $B^+ \to D^+D^-K^+$ process by the LHCb collaboration~\cite{LHCb:2020pxc}. The $X(3915)$ was reported in the $B \to K\omega J/\psi$ decay by the Belle~\cite{Belle:2004lle}, and have also been reported by several experiments~\cite{BaBar:2010wfc,LHCb:2020pxc}.
The $X(4010)$ has been reported in the $1^{++}$ channel in the $B^+ \to D^{*\pm}D^\mp K^+$ decays by the LHCb recently as a new charmonium state~\cite{LHCb:2024vfz}. 
In the case of $J^{PC} = 2^{++}$, the $Z(3930)$ is a possible candidate, that was observed in the $e^+e^- \to e^+e^- D\bar{D}$ process by the Belle~\cite{Belle:2005rte}, and have also been reported by the BaBar and LHCb~\cite{BaBar:2010jfn,LHCb:2019lnr}. 
These states are located near the $D^{(*)}\bar{D}^{(*)}$ thresholds and in the vicinity of the $\chi_{cJ}(2P)$ states ($J=0,1,2$), which are predicted by constituent quark models~\cite{Godfrey:1985xj,Barnes:2005pb}.

In this study, we analyze hidden-charm tetraquark states using the mixture model composed of a $c\bar{c}$ core coupled to a $D^{(*)}\bar{D}^{(*)}$ hadronic molecular component. As the $c\bar{c}$ core, we employ the $\chi_{cJ}(2P)$ states ($J=0,1,2$) predicted by the constituent quark models~\cite{Godfrey:1985xj,Barnes:2005pb}. For the $D^{(*)}\bar{D}^{(*)}$ channels, we consider those near the energy region of the $X$ states, including $D$-wave configurations and $D_s\bar{D}_s$ channels. 
In this paper, we denote the mixture states as the physical $\chi_{cJ}(2P)$, referring to states composed of both $c\bar{c}$ and $D^{(*)}\bar{D}^{(*)}$ components. These are to be distinguished from the bare $\chi_{cJ}(2P)$ states predicted by the constituent quark model as a $c\bar{c}$ charmonium.
We conduct a systematic study focusing on the $X(3860)$, $X(3872)$, and $Z(3930)$, which are treated as bound states under the assumption that the couplings to open channels are negligible within the present framework. This assumption is justified by the fact that the thresholds of the relevant open channels are either significantly separated from the masses of the $X$ and $Z$ states or involve higher partial waves that suppress their contributions. The masses of the exotic states are summarized in Table~\ref{tab:Xmasses}.
Following the approach developed in Refs.~\cite{Takizawa:2012hy,Yamaguchi:2019vea}, we introduce a transition potential between the $c\bar{c}$ and $D^{(*)}\bar{D}^{(*)}$ components.
As a refinement beyond Refs.~\cite{Takizawa:2012hy,Yamaguchi:2019vea}, we incorporate more realistic hadron-hadron interactions between the $D^{(*)}$ and $\bar{D}^{(*)}$ mesons by including pseudoscalar-meson and vector-meson exchange potentials. The free parameters in the model are determined so as to reproduce the observed masses of the $X(3872)$ and $Z(3930)$, and we predict a bound state with $J^{PC} = 0^{++}$ corresponding to the $X(3860)$, which currently has large experimental uncertainties.
We also investigate the internal structure of the $X(3860)$, $X(3872)$, and $Z(3930)$ within the model. 

The structure of this paper is as follows: In Section~\ref{sec:formalism}, we describe the formalism, including the Schr\"odinger equations, meson-exchange potentials, and the transition potential. In Section~\ref{sec:result}, we determine the model parameters and present numerical results obtained by solving the coupled-channel Schr\"odinger equations. Section~\ref{sec:summary} is devoted to the summary.

\begin{table}
	\caption{\label{tab:Xmasses} Masses and $J^{PC}$ of the $X$ states studied in this paper~\cite{ParticleDataGroup:2024cfk}. The values are given in units of MeV. }
	%\begin{ruledtabular}
		\begin{tabular}{c|c|c}
            \hline\hline
			$J^{PC}$ &Particles &  Mass [MeV]                                                                                                                                                                                                 \\
			\hline
			$0^{++}$ & $X(3860)$ $(\chi_{c0}(3860))$& $3862^{+50}_{-35}$                                                      \\
			$1^{++}$ & $X(3872)$ $(\chi_{c1}(3872))$& $3871.64 \pm 0.06$                              \\
			$2^{++}$ & $Z(3930)$ $(\chi_{c2}(3930))$& $3922.5 \pm 1.0$ \\
                        \hline\hline    
		\end{tabular}
	%\end{ruledtabular}
\end{table}
\section{Formalism}\label{sec:formalism}
\subsection{Schrödinger equation}\label{subsec:schrodinger-equation}
We analyze the physical $\chi_{cJ}(2P)$ state as a superposition of the bare $\chi_{cJ}(2P)$ state and hadronic molecules.
For the hadronic molecule component, we consider the {$D^{(*)}\bar{D}^{(*)}$ and $D_s D_s$} channels listed in Table \ref{tab:channels}.
To treat the physical $\chi_{cJ}(2P)$ as bound states, we ignore the $D\bar{D}$ channel for $0^{++}$ and the $D\bar{D}$ and $D\bar{D}^*$ channels for $2^{++}$.
\footnote{{In fact, a mass difference between the $D\bar{D}$ threshold and the bare $\chi_{cJ}(2P)$ state is large as seen later, and thus we consider that contributions from the $D\bar{D}$ channels are small, while couplings between the bare $\chi_{cJ}(2P)$ and $D\bar{D}^*$ and between the bare $\chi_{cJ}(2P)$ and $D^*\bar{D}^*$ are important. For $2^{++}$, not only $D\bar{D}$, but also $D\bar{D}^*$ is ignored, because the $S$-wave state of $D\bar{D}^*$ is absent.}
} 
For simplicity, we also ignore the $G$-wave, as its impact is minimal.
\begin{table*}
	\caption{\label{tab:channels}Channels considered in the hadronic molecule part {for given $J^{PC}$}.}
	\begin{ruledtabular}
		\begin{tabular}{c|c}
			$J^{PC}$ & Channel                                                                                                                                                                                                 \\
			\hline
			$0^{++}$ & $D_s^+D_s^-(^1S_0)$,\quad$D^{*0}\bar{D}^{*0}(^1S_0)$,\quad$D^{*+}D^{*-}(^1S_0)$,\quad$D^{*0}\bar{D}^{*0}(^5D_0)$,\quad$D^{*+}D^{*-}(^5D_0)$                                                             \\
			$1^{++}$ & $[D^0\bar{D}^{*0}](^3S_1)$,\quad$[D^+D^{*-}](^3S_1)$,\quad$[D^0\bar{D}^{*0}](^3D_1)$,\quad$[D^+D^{*-}](^3D_1)$,\quad$D^{*0}\bar{D}^{*0}(^5D_1)$,\quad$D^{*+}D^{*-}(^5D_1)$                              \\
			$2^{++}$ & $D_s^+D_s^-(^1D_2)$,\quad$D^{*0}\bar{D}^{*0}(^5S_2)$,\quad$D^{*+}D^{*-}(^5S_2)$,\quad$D^{*0}\bar{D}^{*0}(^1D_2)$,\quad$D^{*+}D^{*-}(^1D_2)$,\quad$D^{*0}\bar{D}^{*0}(^5D_2)$,\quad$D^{*+}D^{*-}(^5D_2)$
		\end{tabular}
	\end{ruledtabular}
\end{table*}

The coupled-channel Schr\"odinger equation is expressed as follows:
\begin{gather}
	\mathcal{H}\Psi = E\Psi, \label{eq:Schrödinger}\\
	\mathcal{H} =
	\begin{pmatrix}
		H_0+V_\mathrm{OBE} & \mathcal{U}^\dagger                \\
		\mathcal{U}        & m_{\chi_{cJ}}-m_\mathrm{threshold}
	\end{pmatrix}, \label{eq:total_hamiltonian} \\
	\Psi =
	\begin{pmatrix}
		c_1\ket{\mathrm{HM}_1} \\
		c_2\ket{\mathrm{HM}_2} \\
		\vdots                 \\
		c_\chi\ket{\chi_{cJ}(2P)}
	\end{pmatrix} . %,
        \label{eq:wavefunc_HM_chicJ}
\end{gather}
{$H_0+V_\mathrm{OBE}$ in \eqref{eq:total_hamiltonian} is the hadronic molecular component of the total Hamiltonian $\mathcal{H}$,} where $H_0$ is the kinetic term, {and} $V_\mathrm{OBE}$ is the one-boson-exchange (OBE) potential. 
{In this study, we consider $\pi$, $\eta$, $K$, $\rho$, $\omega$, $K^*$, and $\phi$ exchanges as OBE potentials.}
$\mathcal{U}$ {in the off-diagonal component} is the transition potential {between  the hadronic molecular and bare $\chi_{cJ}$ components. } 
$m_{\chi_{cJ}}$ is the mass of the bare $\chi_{cJ}(2P)$, and $m_\mathrm{threshold}$ is the sum of the masses of the threshold particles. 
In \eqref{eq:wavefunc_HM_chicJ}, $\ket{\mathrm{HM}_i}$ is the $i$-th hadronic molecule state, and $\ket{\chi_{cJ}(2P)}$ is the bare $\chi_{cJ}(2P)$ state.

\subsection{Heavy meson chiral Lagrangian}\label{subsec:heavy-meson-chiral-lagrangian}
The $V_\mathrm{OBE}$ is derived from the heavy meson chiral Lagrangian.
Pseudoscalar mesons $D$ and vector mesons $D^*$ in the heavy quark limit ($m_c \rightarrow \infty$) are degenerate by heavy quark spin symmetry~\cite{Neubert:1993mb}.
Therefore, we define the heavy meson field $H_a^{(Q)}$ as
\begin{equation}
	H^{(Q)}_a \equiv \frac{1+\slashed{v}}{2}(P^{*\mu}_a\gamma_\mu-P_a\gamma^5)
\end{equation}
with pseudoscalar heavy meson field $P=(D^0,D^+,D_s^+)$ and vector heavy meson field $P^{*}=(D^{*0},D^{*+},D_s^{*+})$.
Then $v$ is the four-velocity of the heavy quark, and $a$ is the flavor index.
The conjugate field $\bar{H}_a^{(Q)}$ to $H_a^{(Q)}$ is given by
\begin{equation}
	\bar{H}^{(Q)}_a \equiv \gamma^0(H^{(Q)}_a)^\dagger\gamma^0 = (P^{*\mu\dagger}_a\gamma_\mu+P^\dagger_a\gamma^5)\frac{1+\slashed{v}}{2}.
\end{equation}
The anti-particle field of the heavy meson is defined as
\begin{gather}
	\begin{split}
		H^{(\bar{Q})}_a & \equiv C(\mathcal{C}H^{(Q)}_a\mathcal{C}^{-1})^{\top}C^{-1}              \\
		                & = (\bar{P}^{*\mu}_a\gamma_\mu-\bar{P}_a\gamma^5)\frac{1-\slashed{v}}{2},
	\end{split}\\
	\bar{H}^{(\bar{Q})}_a \equiv \gamma^0(H^{(\bar{Q})}_a)^\dagger\gamma^0 = \frac{1-\slashed{v}}{2}(\bar{P}^{*\mu\dagger}_a\gamma_\mu+\bar{P}^\dagger_a\gamma^5),
\end{gather}
taking charge conjugation.

We construct the interaction Lagrangian involving the light pseudoscalar mesons to satisfy heavy quark spin symmetry, chiral symmetry, and Lorentz symmetry~\cite{Wise:1992hn,Casalbuoni:1996pg}:
\begin{equation}
	\mathcal{L}_{MHH} = ig\Tr{\left[H^{(Q)}_b\gamma^\mu\gamma^5A_{ba\mu}\bar{H}^{(Q)}_a\right]} + ig\Tr{\left[\bar{H}^{(\bar{Q})}_a\gamma^\mu\gamma^5A_{ab\mu}H^{(\bar{Q})}_b\right]}.
\end{equation}
Here, $g$ is the coupling constant, and $A_{ab\mu}$ is the axial current of the pseudoscalar meson which is given by
\begin{equation}
	A_{\mu} = \frac{1}{2}{\left(\xi^\dagger\partial_\mu\xi-\xi\partial_\mu\xi^\dagger\right)},\quad \xi = \exp{\left(\frac{\bm{\pi}\cdot\bm{\tau}}{2f_\pi}\right)},
\end{equation}
where $f_\pi\simeq\qty{93}{MeV}$ is the pion decay constant, and $\bm{\pi}$ is the pseudoscalar field written as
\begin{equation}
	\frac{\bm{\pi}\cdot\bm{\tau}}{\sqrt{2}} =
	\begin{pmatrix}
		\frac{\pi^0}{\sqrt{2}}+\frac{\eta}{\sqrt{6}} & \pi^+                                         & K^+                     \\
		\pi^-                                        & -\frac{\pi^0}{\sqrt{2}}+\frac{\eta}{\sqrt{6}} & K^0                     \\
		K^-                                          & \bar{K}^0                                     & -\frac{2\eta}{\sqrt{6}}
	\end{pmatrix}.
\end{equation}
We also construct the interaction Lagrangian with light vector mesons using hidden local symmetry~\cite{Bando:1987br,Harada:2003jx,Casalbuoni:1996pg}:
\begin{widetext}
	\begin{multline}
		\mathcal{L}_{VHH} = i\beta\Tr{\left[H^{(Q)}_b v_\mu(V^\mu_{ba}-\rho^\mu_{ba})\bar{H}^{(Q)}_a\right]}
		+i\lambda\Tr{\left[H^{(Q)}_b\sigma_{\mu\nu}F^{\mu\nu}_{ba}\bar{H}^{(Q)}_a\right]}                    \\
		-i\beta\Tr{\left[\bar{H}^{(\bar{Q})}_a v_\mu(V^\mu_{ab}-\rho^\mu_{ab})H^{(\bar{Q})}_b\right]}
		+i\lambda\Tr{\left[\bar{H}^{(\bar{Q})}_a\sigma_{\mu\nu}F^{\prime\mu\nu}_{ab}H^{(\bar{Q})}_b\right]},
	\end{multline}
\end{widetext}
where
\begin{gather}
	V^\mu = \frac{1}{2}{\left(\xi^\dagger\partial^\mu\xi+\xi\partial^\mu\xi^\dagger\right)},\\
	F^{\mu\nu} =\partial^\mu\rho^\nu-\partial^\nu\rho^\mu+[\rho^\mu,\rho^\nu],\\
	F^{\prime\mu\nu} = \partial^\mu\rho^\nu-\partial^\nu\rho^\mu-[\rho^\mu,\rho^\nu], \\
	\rho^\mu = \frac{ig_v}{\sqrt{2}}
	\begin{pmatrix}
		\frac{\rho^0}{\sqrt{2}}+\frac{\omega}{\sqrt{2}} & \rho^+                                           & K^{*+} \\
		\rho^-                                          & -\frac{\rho^0}{\sqrt{2}}+\frac{\omega}{\sqrt{2}} & K^{*0} \\
		K^{*-}                                          & \bar{K}^{*0}                                     & \phi
	\end{pmatrix}^\mu \\
	g_v = \frac{m_\rho}{\sqrt{2}f_\pi},
\end{gather}
and $\beta$ and $\lambda$ are the coupling constants for the vector mesons.

\subsection{One-boson-exchange potential}\label{subsec:one-boson-exchange-potential}
The OBE potential $V_\mathrm{OBE}$ is derived from these heavy meson chiral Lagrangians~\cite{Ohkoda:2012hv}.
For simplicity, we adopt a static approximation that ignores energy transfer.
Terms derived from the delta function are ignored in order to focus on the medium- and long-range interactions.
In the case of $0^{++}$, the OBE potential is given by
	\begin{gather}
		V_\mathrm{OBE}  = V_\pi+V_\eta+V_{\lambda}+V_{\beta}, \\
		V_\pi        = \frac{1}{3}{\left(\frac{g}{2f_\pi}\right)}^2
		\begin{pmatrix}
			0             & 2\sqrt{3}C_K    & 2\sqrt{3}C_K    & -2\sqrt{6}T_K   & -2\sqrt{6}T_K   \\
			2\sqrt{3}C_K  & -2C_\pi         & -4C_\pi         & -\sqrt{2}T_\pi  & -2\sqrt{2}T_\pi \\
			2\sqrt{3}C_K  & -4C_\pi         & -2C_\pi         & -2\sqrt{2}T_\pi & -\sqrt{2}T_\pi  \\
			-2\sqrt{6}T_K & -\sqrt{2}T_\pi  & -2\sqrt{2}T_\pi & C_\pi-2T_\pi    & 2C_\pi-4T_\pi   \\
			-2\sqrt{6}T_K & -2\sqrt{2}T_\pi & -\sqrt{2}T_\pi  & 2C_\pi-4T_\pi   & C_\pi-2T_\pi
		\end{pmatrix},                                                                         
        \end{gather}
        \begin{gather}
		V_\eta       = \frac{1}{9}{\left(\frac{g}{2f_\pi}\right)}^2
		\begin{pmatrix}
			0 & 0               & 0               & 0               & 0               \\
			0 & -2C_\eta        & 0               & -\sqrt{2}T_\eta & 0               \\
			0 & 0               & -2C_\eta        & 0               & -\sqrt{2}T_\eta \\
			0 & -\sqrt{2}T_\eta & 0               & C_\eta-2T_\eta  & 0               \\
			0 & 0               & -\sqrt{2}T_\eta & 0               & C_\eta-2T_\eta
		\end{pmatrix}, \\
		V_{\lambda}  = \frac{1}{3}{\left(\lambda g_v\right)}^2
		\begin{pmatrix}
			0                & 4\sqrt{3}C_{K^{*}}                & 4\sqrt{3}C_{K^{*}}                & 2\sqrt{6}T_{K^{*}}                    & 2\sqrt{6}T_{K^{*}}                    \\
			4\sqrt{3}C_{K^{*}} & -4C_\rho-4C_\omega              & -8C_\rho                        & \sqrt{2}T_\rho+\sqrt{2}T_\omega     & 2\sqrt{2}T_\rho                     \\
			4\sqrt{3}C_{K^{*}} & -8C_\rho                        & -4C_\rho-4C_\omega              & 2\sqrt{2}T_\rho                     & \sqrt{2}T_\rho+\sqrt{2}T_\omega     \\
			2\sqrt{6}T_{K^{*}} & \sqrt{2}T_\rho+\sqrt{2}T_\omega & 2\sqrt{2}T_\rho                 & 2C_\rho+2T_\rho+2C_\omega+2T_\omega & 4C_\rho+4T_\rho                     \\
			2\sqrt{6}T_{K^{*}} & 2\sqrt{2}T_\rho                 & \sqrt{2}T_\rho+\sqrt{2}T_\omega & 4C_\rho+4T_\rho                     & 2C_\rho+2T_\rho+2C_\omega+2T_\omega
		\end{pmatrix},         \\  
		V_{\beta}    = {\left(\frac{\beta g_v}{2}\right)}^2
		\begin{pmatrix}
			-2C_\phi/m_\phi^2 & 0                                    & 0                                    & 0                                    & 0                                    \\
			0                 & -C_\rho/m_\rho^2-C_\omega/m_\omega^2 & -2C_\rho/m_\rho^2                    & 0                                    & 0                                    \\
			0                 & -2C_\rho/m_\rho^2                    & -C_\rho/m_\rho^2-C_\omega/m_\omega^2 & 0                                    & 0                                    \\
			0                 & 0                                    & 0                                    & -C_\rho/m_\rho^2-C_\omega/m_\omega^2 & -2C_\rho/m_\rho^2                    \\
			0                 & 0                                    & 0                                    & -2C_\rho/m_\rho^2                    & -C_\rho/m_\rho^2-C_\omega/m_\omega^2
		\end{pmatrix},            
        \end{gather}
where $C_\mathrm{meson}$ and $T_\mathrm{meson}$ are the central and tensor potentials given by
\begin{gather}
C_\mathrm{meson} = \frac{{m_\mathrm{meson}}^2}{4\pi}{\left[\frac{e^{-m_\mathrm{meson}r}}{r}-\frac{e^{-\Lambda_\mathrm{meson}r}}{r}-\frac{{\Lambda_\mathrm{meson}}^2-{m_\mathrm{meson}}^2}{2\Lambda_\mathrm{meson}}e^{-\Lambda_\mathrm{meson}r}\right]}, \\
		\begin{multlined}
			T_\mathrm{meson} = \frac{1}{4\pi}(3+3m_\mathrm{meson}r+{m_\mathrm{meson}}^2r^2)\frac{e^{-m_\mathrm{meson}r}}{r^3}-\frac{1}{4\pi}(3+3\Lambda_\mathrm{meson}r+{\Lambda_\mathrm{meson}}^2r^2)\frac{e^{-\Lambda_\mathrm{meson}r}}{r^3}\\
			-\frac{1}{4\pi}\frac{{\Lambda_\mathrm{meson}}^2-{m_\mathrm{meson}}^2}{2}(1+\Lambda_\mathrm{meson}r)\frac{e^{-\Lambda_\mathrm{meson}r}}{r},
    \end{multlined}
\end{gather}
 respectively. For $1^{++}$ and $2^{++}$, see Appendix~\ref{apd:obe}.
 
{To derive the potentials, we introduce the form factor
\begin{equation}
    F(\vec{q},m_\mathrm{meson}) = \left(\frac{\Lambda_\mathrm{meson} - m_\mathrm{meson}}{\Lambda_\mathrm{meson} + \vec{q}^2}\right)^2 ,
\end{equation}
in the momentum space.
We employ the cutoff parameter depending on the exchanged-meson mass as $\Lambda_\mathrm{meson} = \qty{220}{MeV}\cdot\alpha+m_\mathrm{meson}$. This parametrization of the cutoff was introduced in Ref.~\cite{Cheng:2004ru}, where $\alpha=1.1$ was chosen in the $B\to D\pi$ decay analysis. In this study, we use $\alpha=0.7$, $1.0$ and $1.3$, and see the dependence of the results on the cutoff as a function of $\alpha$.
}

\subsection{Transition potential}\label{subsec:transition-potential}
The transition potential $\mathcal{U}$ describes the interaction between the bare $\chi_{cJ}(2P)$ state and the hadronic molecule.
We consider the bare $\chi_{cJ}(2P)$ couples to the $S$-wave hadronic molecule.
For instance in the case of $1^{++}$, $\mathcal{U}=(U,U,0,0,0,0,0)$.
The transition potential is given by~\cite{Takizawa:2012hy}
\begin{equation}
	\braket[3]{\chi_{cJ}(2P)}{U}{HM} = \int d^3x\,\braket[3]{\chi_{cJ}(2P)}{U}{\bm{x}}\braket{\bm{x}}{HM} = \int d^3x\,\sqrt{2}\pi f_\mathrm{spin}g_{c\bar{c}}\Lambda_q^\frac{3}{2}\frac{e^{-\Lambda_q r}}{r}Y^l_m(\Omega)\braket{\bm{x}}{HM},
\end{equation}
where $f_\mathrm{spin}$ is the spin factor {determined by the Clebsch-Gordan coefficients}, $g_{c\bar{c}}$ is the coupling constant, $\Lambda_q$ is a parameter that determines the spread of $\chi_{cJ}(2P)$ and $U$, $Y^l_m(\Omega)$ is the spherical harmonics, and $\braket{\bm{x}}{HM}$ is the wave function of the hadronic molecule. $g_{c\bar{c}}$ and $\Lambda_q$ are free parameters.
We use the $f_\mathrm{spin}$ values in Table~\ref{tab:spin-factor}.
\begin{table}
	\caption{\label{tab:spin-factor}Spin factor $f_\mathrm{spin}$ for each channel.}
%	\begin{ruledtabular}
		\begin{tabular}{ccccc} %{c|c|c|c|c}
            \hline\hline
			                  & $D_s^+D_s^-(^1S_0)$ & $D^{*}\bar{D}^{*}(^1S_0)$ & $D\bar{D}^{*}(^3S_1)$ & $D^{*}\bar{D}^{*}(^5S_2)$ \\
			\hline
			$f_\mathrm{spin}$ & $\frac{1}{2}$       & $\frac{\sqrt{3}}{2}$      & $1$                   & $1$                       \\
                        \hline\hline    
		\end{tabular}
%	\end{ruledtabular}
\end{table}

\section{Numerical Results}\label{sec:result}
{In this section, we study bound states of the physical $\chi_{cJ}(2P)$ states.} 
{The coupling constants of the OBE potentials are determined by experimental data and Lattice QCD resutls, which are summarized in Table~\ref{tab:parameters}.}
The parameter $g$ is determined from the experimental decay width of $D^{*+}\rightarrow D^+\pi^0$~\cite{ParticleDataGroup:2024cfk}.
The parameter $\beta$ is determined from the lattice calculation~\cite{Detmold:2007wk} and $\lambda$ is estimated using the form factor in $B$ decay~\cite{Isola:2003fh}.
{As the masses of the bare $\chi_{cJ}(2P)$ state, $m_{\chi_{cJ}}$, we employ the values predicted by the Godfrey-Isgur (GI) model~\cite{Godfrey:1985xj,Barnes:2005pb}. }
\begin{table}
    \caption{\label{tab:chi_mass}The masses of the bare $\chi_{cJ}(2P)$ state}
%    	\begin{ruledtabular}
    \begin{tabular}{ccc}
        \hline\hline
         $\chi_{c0}$&$\chi_{c1}$&$\chi_{c2}$\\
         \hline
         \qty{3916}{MeV} & \qty{3953}{MeV} & \qty{3979}{MeV} \\
                 \hline\hline   
    \end{tabular}
%    \end{ruledtabular}
\end{table}

{In the model constructed in the previous section, we have free parameters that are $\alpha$, $g_{c\bar{c}}$, and $\Lambda_q$. To fix them, we use the experimental mass spectra of the hidden-charm tetraquarks. In this study, we focus on $X(3860)$ with $J^{PC}=0^{++}$, $X(3872)$ with $J^{PC}=1^{++}$, $Z(3930)$ with $J^{PC}=2^{++}$. Since the $D\bar{D}$ channels are ignored, these tetraquarks can be regarded as a bound state in our model space. }
While $X(3860)$ {has been} reported only in the Belle experiment~\cite{Belle:2017egg} {and the data has large ambiguity}, $X(3872)$ and $Z(3930)$ { have been done } in several experiments~\cite{Belle:2003nnu,CDF:2003cab,D0:2004zmu,BaBar:2004oro,LHCb:2011zzp,CMS:2013fpt,BESIII:2013fnz,ATLAS:2016kwu,Belle:2005rte,BaBar:2010jfn,LHCb:2019lnr}.
Therefore, we use the central masses of $X(3872)$ and $Z(3930)$ as input. % in this paper.
The three free parameters, $\alpha$, $g_{c\bar{c}}$, and $\Lambda_q$, are determined using two inputs.
{Thus, first, we fix $\alpha$ as $\alpha=0.7$, $1.0$ or $1.3$. For each $\alpha$, we determine the values of $g_{c\bar{c}}$ and $\Lambda_q$ to reproduce the masses of $X(3872)$ and $Z(3930)$. 
We use the Gaussian expansion method~\cite{Hiyama:2003cu} to solve the equation in \eqref{eq:Schrödinger} and obtain the mass of $X(3872)$ and $Z(3930)$ as the physical $\chi_{cJ}(2P)$ for $J^{PC}=1^{++}$ and $2^{++}$, respectively.
The values for each $\alpha$, obtained as three sets of parameters, are summarized in Table~\ref{tab:free-parameters}. By using the parameters, we compute the mass of the $0^{++}$ bound state that may correspond to $X(3860)$.}
\begin{table}
	\caption{\label{tab:free-parameters} {Parameters $g_{c\bar{c}}$ and $\Lambda_q$ determined to reproduce the masses of $X(3872)$ and $Z(3930)$ for given $\alpha=0.7$, $1.0$ and $1.3$, respectively.}}
%	\begin{ruledtabular}
		\begin{tabular}{cccc}
            \hline\hline
			$\alpha$                & $0.7$    & $1.0$    & $1.3$    \\
			$g_{c\bar{c}}$          & $0.0448$ & $0.0427$ & $0.0409$ \\
			$\Lambda_q$(\unit{MeV}) & $2260$   & $3089$   & $4647$   \\
                        \hline\hline    
		\end{tabular}
%	\end{ruledtabular}
\end{table}

\begin{table}
	\caption{\label{tab:parameters} {Coupling constants of the OBE potentials.}}
%	\begin{ruledtabular}
		\begin{tabular}{ccc}
            \hline\hline
			$g$    & $\beta$ & $\lambda$            \\
			\hline
			$0.55$ & $0.9$   & \qty{0.56}{GeV^{-1}} \\
                        \hline\hline    
		\end{tabular}
%	\end{ruledtabular}
\end{table}

\begin{figure}
	\centering
	\includegraphics[width=0.6\textwidth]{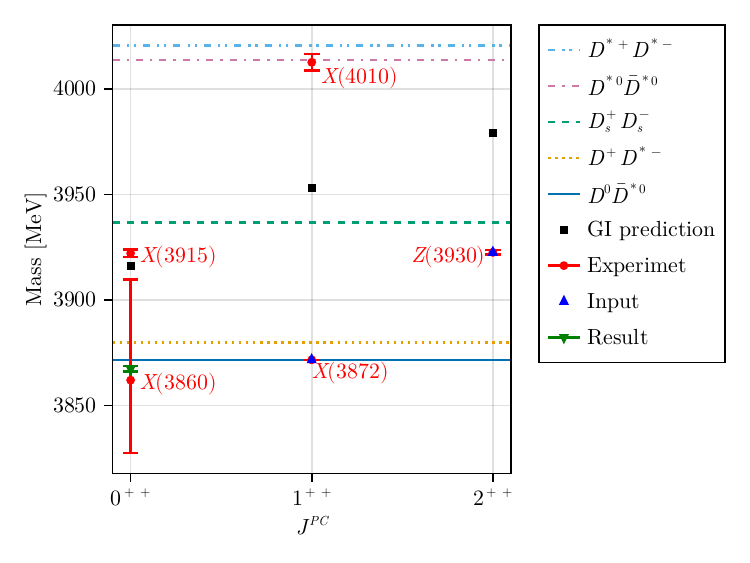}
	\caption{\label{fig:mass-of-x(3860)}Calculated mass of bound state of $0^{++}$ at each $\alpha$. {The red dots with an error are the empirical mass spectra of the tetraquarks. The blue triangles are the computed masses reproducing the masses of $X(3872)$ and $Z(3930)$. The green inverted triangle with an error is the computed mass of the $0^{++}$ bound state, where the error indicates the distribution of the obtained masses for each $\alpha$.   The horizontal lines show threshold energies of $D^{(*)}_s\bar{D}^{(*)}_s$. The solid squares are mass spectra of the bare $\chi_{cJ}(2P)$, $m_{\chi_{cJ}}$, predicted by the GI model~\cite{Godfrey:1985xj,Barnes:2005pb}. }}
\end{figure}

{Using the obtained model, we predict the $0^{++}$ bound state.}
The calculated masses for each $\alpha$ are shown in Table~\ref{tab:mass-of-x(3860)} and plotted in Figure~\ref{fig:mass-of-x(3860)}.
{The obtained $0^{++}$ bound state would correspond to $X(3860)$. There have been  discussions about the existence of $X(3860)$, where it has been reported by Belle~\cite{Belle:2017egg}, while no evidence of $X(3860)$  has been found by LHCb~\cite{LHCb:2020pxc}. }
{It is interesting that we obtain the $0^{++}$ bound state by using the model constructed to reproduce $X(3872)$ and $Z(3930)$.}
These calculated masses are consistent with the {Belle data} for $X(3860)$.
{Table~\ref{tab:mass-of-x(3860)} shows the low $\alpha$ dependence of the results.}
\begin{table}
	\caption{\label{tab:mass-of-x(3860)}Calculated mass of bound state of $0^{++}$ at each $\alpha$}
%	\begin{ruledtabular}
		\begin{tabular}{cccc}
            \hline\hline
			$\alpha$           & $0.7$     & $1.0$     & $1.3$     \\
			Mass\,(\unit{MeV}) & $3868.62$ & $3867.31$ & $3866.07$ \\
                        \hline\hline    
		\end{tabular}
%	\end{ruledtabular}
\end{table}
The mixing ratios {of the} $0^{++}$ {bound state} are shown in Table~\ref{tab:mixing-ratio-0}, and the corresponding wave function is depicted in Figure~\ref{fig:wave-function-0}.
The bare $\chi_{c0}(2P)$ component exceeds \qty{90}{\percent}, making it the dominant component, {while the contributions from the hadronic molecular components are small. In fact, the bound state is obtained below the bare $\chi_{c0}(2P)$\footnote{We note that the $D\bar{D}^*$ channel close to the obtained bound state is absent for $J^{PC}=0^{++}$. See Table~\ref{tab:channels}.}.} The differences in the mixing ratios of hadronic molecules are primarily due to $f_\mathrm{spin}$.
The mixing ratio of the $D$-wave is \qty{0}{\percent} because we do not consider its coupling with the bare $\chi_{c0}(2P)$ state.
In Table~\ref{tab:expectation-value-0}, the expectation values of the kinetic term and potential terms for the $0^{++}$ bound state are summarized. We find that the strong attraction is produced by the transition potential $U$ whose contribution is larger than the kinetic-term one, and thus this potential dominates to make the bound state, while the meson exchange potentials in the hadronic molecules play a minor role.

{We also discuss the properties of the bound states for $J^{PC}=1^{++}$ and $2^{++}$, corresponding to $X(3872)$ and $Z(3930)$, that are used as input.}
Table~\ref{tab:mixing-ratio-1} shows the mixing ratios for $1^{++}$, and Figure~\ref{fig:wave-function-1} illustrates the corresponding wave function.
Since the mass of $X(3872)$ is very close to the threshold of $D^{0}\bar{D}^{*0}({}^3S_1)$, the dominant component is $D^{0}\bar{D}^{*0}({}^3S_1)$, {while the $D^{+}{D}^{*-}({}^3S_1)$ component is suppressed. It leads} to the breaking of isospin symmetry.
The bare $\chi_{c1}(2P)$ is the second dominant term, {and this mixing ratio is larger than the one obtained in our previous works~\cite{Yamaguchi:2019vea}.
}
This is because the coupling with the bare $\chi_{cJ}(2P)$ {has} to be large to simultaneously reproduce the mass of $Z(3930)$, {where the transition potential plays an important role as seen later. 
The expectation values of the kinetic and potential terms of the $1^{++}$ bound state are summarized in Table~\ref{tab:expectation-value-1}. In this case, we also obtain the strong attraction from the transition potential as seen in the case for $0^{++}$. However, the magnitude is smaller than that of the kinetic term. Thus, the attraction from the transition potential is not enough to produce the bound state, while the attraction from the meson exchange potentials is also important. The expectation values also show the difference between structures of the $0^{++}$ and $1^{++}$ bound states, where the bare $\chi_{cJ}(2P)$ component is dominant for $0^{++}$, while the molecular component is for $1^{++}.$

The mixing ratios for $2^{++}$ are presented in Table~\ref{tab:mixing-ratio-2}, and the wave function is shown in Figure~\ref{fig:wave-function-2}.
The results for the $2^{++}$ case are very similar to those for the $0^{++}$ case, {where the bare $\chi_{c2}(2P)$ state is a main component. } The primary difference {is} that the mixing ratio of $D_s^+D_s^-$ is smaller for the $2^{++}$ case, {while the $D_s^+D_s^-$ threshold energy is most close to the mass of $Z(3930)$}.
This is because the coupling between the bare $\chi_{cJ}(2P)$ and the $D$-wave {$D_s^+D_s^-$ state is absent.}
The expectation values of the kinetic and potential terms of the $2^{++}$ bound state are summarized in Table~\ref{tab:expectation-value-2}. As seen in the case of the $0^{++}$ bound state, the strong attraction of the transition potential whose magnitude is larger than that of the kinetic term is obtained.

\begin{table}[H]
	\caption{\label{tab:mixing-ratio-0}Mixing ratio of {the $0^{++}$ bound state for each $\alpha$. Mass shows the corresponding threshold energies.}}
	\begin{ruledtabular}
		\begin{tabular}{ccccccc}
			             & $D_s^+{D}_s^{-}({}^1S_0)$ & $D^{*0}\bar{D}^{*0}({}^1S_0)$ & $D^{*+}D^{*-}({}^1S_0)$ & $D^{*0}\bar{D}^{*0}({}^5D_0)$ & $D^{*+}D^{*-}({}^5D_0)$ & bare $\chi_{c0}(2P)$  \\
			Mass\,(MeV)  & $3937$                    & $4014$                        & $4021$                  & $4014$                        & $4021$                  & $3916$                \\
			\hline
			$\alpha=0.7$ & \qty{1.11}{\percent}      & \qty{2.81}{\percent}          & \qty{2.71}{\percent}    & \qty{0.00}{\percent}          & \qty{0.00}{\percent}    & \qty{93.37}{\percent} \\
			$\alpha=1.0$ & \qty{0.66}{\percent}      & \qty{2.53}{\percent}          & \qty{2.45}{\percent}    & \qty{0.00}{\percent}          & \qty{0.00}{\percent}    & \qty{94.36}{\percent} \\
			$\alpha=1.3$ & \qty{0.32}{\percent}      & \qty{2.16}{\percent}          & \qty{2.10}{\percent}    & \qty{0.00}{\percent}          & \qty{0.00}{\percent}    & \qty{95.42}{\percent}
		\end{tabular}
	\end{ruledtabular}
\end{table}
\begin{table}[H]
	\caption{\label{tab:mixing-ratio-1}Mixing ratio of {the $1^{++}$ bound state for each $\alpha$. Mass shows the corresponding threshold energies.}}
	\begin{ruledtabular}
		\begin{tabular}{cccccccc}
			             & $D^{0}\bar{D}^{*0}({}^3S_1)$ & $D^+D^{*-}(^3S_1)$   & $D^{0}\bar{D}^{*0}(^3D_1)$ & $D^+D^{*-}(^3D_1)$   & $D^{*0}\bar{D}^{*0}(^5D_1)$ & $D^{*+}{D}^{*-}(^5D_1)$ & bare  $\chi_{c1}(2P)$ \\
			Mass\,(MeV)  & $3872$                       & $3880$               & $3872$                     & $3880$               & $4014$                      & $4021$                  & $3953$                \\
			\hline
			$\alpha=0.7$ & \qty{80.49}{\percent}        & \qty{5.05}{\percent} & \qty{0.00}{\percent}       & \qty{0.00}{\percent} & \qty{0.00}{\percent}        & \qty{0.00}{\percent}    & \qty{14.46}{\percent} \\
			$\alpha=1.0$ & \qty{82.03}{\percent}        & \qty{4.97}{\percent} & \qty{0.01}{\percent}       & \qty{0.01}{\percent} & \qty{0.00}{\percent}        & \qty{0.00}{\percent}    & \qty{12.99}{\percent} \\
			$\alpha=1.3$ & \qty{84.76}{\percent}        & \qty{4.86}{\percent} & \qty{0.01}{\percent}       & \qty{0.01}{\percent} & \qty{0.01}{\percent}        & \qty{0.01}{\percent}    & \qty{10.33}{\percent} \\
		\end{tabular}
	\end{ruledtabular}
\end{table}
\begin{table}[H]
	\caption{\label{tab:mixing-ratio-2}Mixing ratio of {the $2^{++}$ bound state for each $\alpha$. Mass shows the corresponding threshold energies.}}
	\begin{ruledtabular}
		\begin{tabular}{cccccccccc}
			             & $D_s^+{D}_s^{-}({}^1D_2)$ & $D^{*0}\bar{D}^{*0}({}^5S_2)$ & $D^{*+}{D}^{*-}({}^5S_2)$ & $D^{*0}\bar{D}^{*0}({}^1D_2)$ & $D^{*+}{D}^{*-}({}^1D_2)$ & $D^{*0}\bar{D}^{*0}({}^5D_2)$ & $D^{*+}{D}^{*-}({}^5D_2)$ & bare $\chi_{c2}(2P)$  \\
			Mass\,(MeV)  & $3937$                    & $4014$                        & $4021$                    & $4014$                        & $4021$                    & $4014$                        & $4021$                    & $3979$                \\
			\hline
			$\alpha=0.7$ & \qty{0.00}{\percent}      & \qty{4.61}{\percent}          & \qty{4.38}{\percent}      & \qty{0.00}{\percent}          & \qty{0.00}{\percent}      & \qty{0.00}{\percent}          & \qty{0.00}{\percent}      & \qty{91.00}{\percent} \\
			$\alpha=1.0$ & \qty{0.00}{\percent}      & \qty{3.67}{\percent}          & \qty{3.50}{\percent}      & \qty{0.00}{\percent}          & \qty{0.00}{\percent}      & \qty{0.00}{\percent}          & \qty{0.00}{\percent}      & \qty{92.83}{\percent} \\
			$\alpha=1.3$ & \qty{0.00}{\percent}      & \qty{2.67}{\percent}          & \qty{2.55}{\percent}      & \qty{0.00}{\percent}          & \qty{0.00}{\percent}      & \qty{0.00}{\percent}          & \qty{0.00}{\percent}      & \qty{94.78}{\percent} \\
		\end{tabular}
	\end{ruledtabular}
\end{table}

\begin{table}[H]
	\caption{\label{tab:expectation-value-0}Expectation values of {the $0^{++}$ bound state in units of MeV. $\langle T\rangle$ stands for the sum of the kinetic and mass terms in the Hamiltonian~\eqref{eq:total_hamiltonian}. $\langle V_i\rangle$ $(i=\pi,\eta,\rho,\omega,\phi)$ is for the meson exchange potential, where the values of the vector meson exchanges are given by summing up $V_\lambda$ and $V_\beta$ terms for each vector mesons. $\langle U\rangle$ is for the transition potential between the hadronic molecule and the bare $\chi_{cJ}(2P)$.} }
	\begin{ruledtabular}
		\begin{tabular}{cccccccc}
			             & $\langle T\rangle$ %$\langle H_0\rangle$ 
                            & $\langle V_\pi\rangle$ & $\langle V_\eta\rangle$ & $\langle V_\rho\rangle$ & $\langle V_\omega\rangle$ & $\langle V_\phi\rangle$ & $\langle U\rangle$ \\
			\hline
			$\alpha=0.7$ & $21.90$            & $0.13$                 & $-0.03$                 & $-0.98$                 & $-0.56$                   & $-0.06$                 & $-88.48$           \\
			$\alpha=1.0$ & $25.31$            & $0.18$                 & $-0.05$                 & $-1.88$                 & $-0.97$                   & $-0.08$                 & $-91.89$           \\
			$\alpha=1.3$ & $28.79$            & $0.17$                 & $-0.07$                 & $-2.83$                 & $-1.32$                   & $-0.07$                 & $-95.28$
		\end{tabular}
	\end{ruledtabular}
\end{table}
\begin{table}[H]
	\caption{\label{tab:expectation-value-1}Expectation value of {the $1^{++}$} bound state. The same convention as Table~\ref{tab:expectation-value-0} is used.}
	\begin{ruledtabular}
		\begin{tabular}{ccccccc}
			             & $\langle T\rangle$  %$\langle H_0\rangle$ 
                            & $\langle V_\pi\rangle$ & $\langle V_\eta\rangle$ & $\langle V_\rho\rangle$ & $\langle V_\omega\rangle$ & $\langle U\rangle$ \\
			\hline
			$\alpha=0.7$ & $25.06$            & $0.06$                 & $0.02$                  & $-1.24$                 & $-0.42$                   & $-23.53$           \\
			$\alpha=1.0$ & $23.85$            & $0.07$                 & $0.03$                  & $-2.16$                 & $-0.73$                   & $-21.12$           \\
			$\alpha=1.3$ & $20.70$            & $0.05$                 & $0.04$                  & $-3.00$                 & $-1.01$                   & $-16.82$
		\end{tabular}
	\end{ruledtabular}
\end{table}
\begin{table}[H]
	\caption{\label{tab:expectation-value-2}Expectation value of {the $2^{++}$ bound state. The same convention as Table~\ref{tab:expectation-value-0} is used.} }
	\begin{ruledtabular}
		\begin{tabular}{cccccccc}
			             & $\langle T\rangle$ %$\langle H_0\rangle$ 
                            & $\langle V_\pi\rangle$ & $\langle V_\eta\rangle$ & $\langle V_\rho\rangle$ & $\langle V_\omega\rangle$ & $\langle V_\phi\rangle$ & $\langle U\rangle$ \\
			\hline
			$\alpha=0.7$ & $89.15$            & $0.05$                 & $0.02$                  & $-0.45$                 & $-0.14$                   & $0.00$                  & $-102.84$           \\
			$\alpha=1.0$ & $91.52$            & $0.06$                 & $0.04$                  & $-0.70$                 & $-0.22$                   & $0.00$                  & $-104.89$           \\
			$\alpha=1.3$ & $93.90$            & $0.06$                 & $0.04$                  & $-0.84$                 & $-0.26$                   & $0.00$                  & $-107.11$
		\end{tabular}
	\end{ruledtabular}
\end{table}

\begin{figure}[h]
    \centering
    \includegraphics[width=0.5\linewidth]{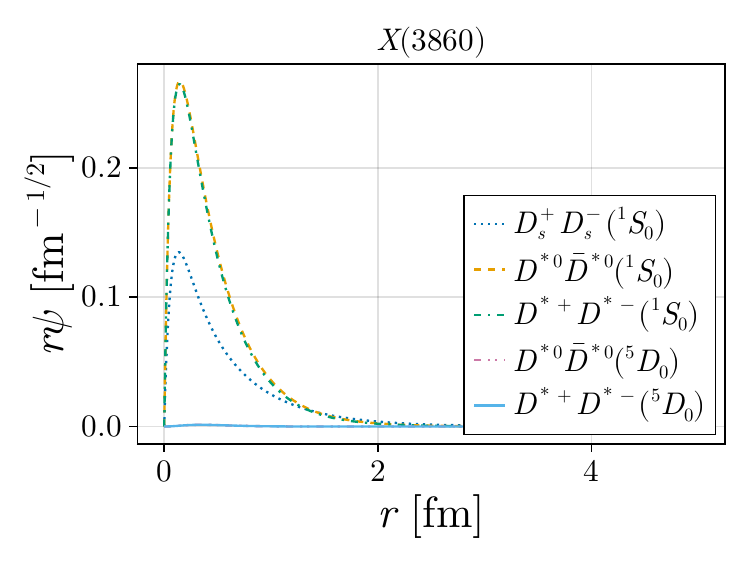}
    \caption{\label{fig:wave-function-0}Wave functions of the $0^{++}$ bound state corresponding to $X(3860)$.}   
\end{figure}

\begin{figure}[h]
    \centering
    \includegraphics[width=0.5\linewidth]{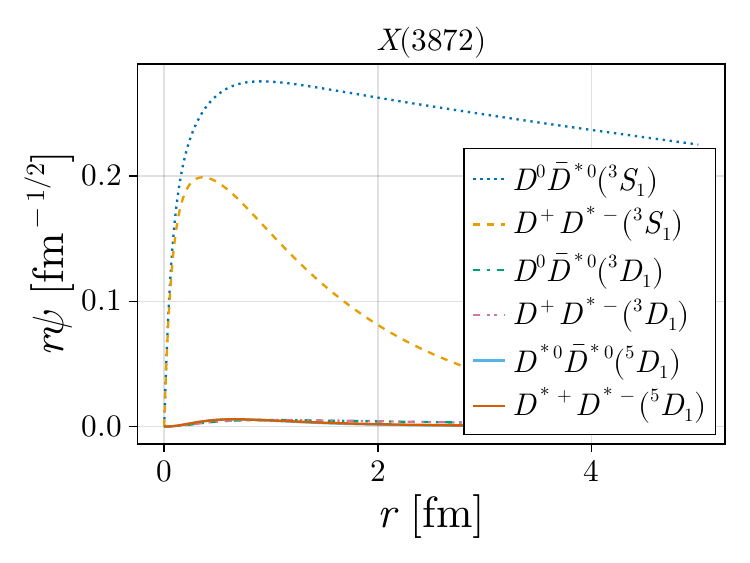}
	\caption{\label{fig:wave-function-1}Wave functions of the $1^{++}$ bound state corresponding to $X(3872)$.}       
\end{figure}

\begin{figure}[h]
    \centering
    \includegraphics[width=0.5\linewidth]{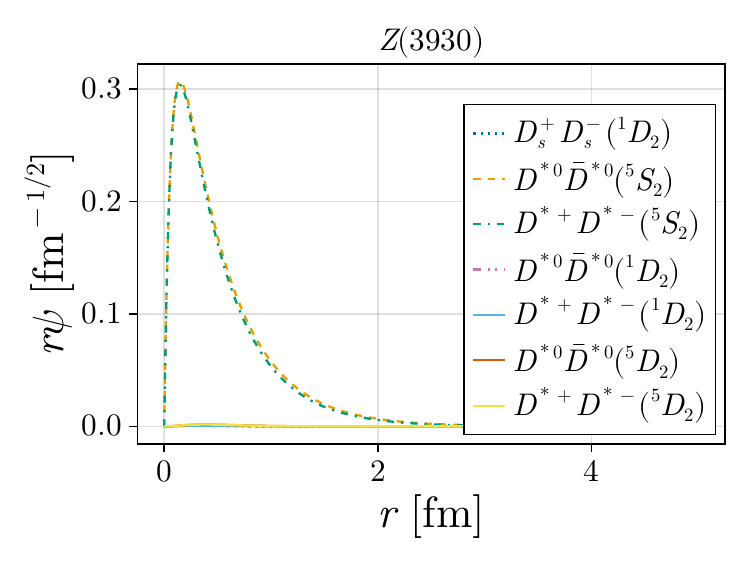}
	\caption{\label{fig:wave-function-2}Wave functions of the $2^{++}$ bound state corresponding to $Z(3930)$.}    
\end{figure}

\section{Summary}\label{sec:summary}

Exotic hadrons are currently a major topic of research in hadron physics. In particular, hidden-charm tetraquark states $XYZ$ have attracted significant attention since the discovery of the $X(3872)$ in 2003, and many related states have been reported in experiments.
In this work, we have performed a systematic analysis of the $X(3872)$ with $J^{PC}=1^{++}$ and its spin partners, the $X(3860)$ with $J^{PC}=0^{++}$ and the $Z(3930)$ with $J^{PC}=2^{++}$, based on a mixture model that incorporates both $c\bar{c}$ states and $D^{(*)}\bar{D}^{(*)}$ hadronic molecular components. The $c\bar{c}$ components are described using the $\chi_{cJ}(2P)$ ($J=0,1,2$) states predicted by the constituent quark model.
The meson-meson interactions are modeled using an effective Lagrangian that respects heavy quark spin and chiral symmetries, incorporating both pseudoscalar and vector meson exchange potentials. The model parameters were determined to reproduce the observed masses of $X(3872)$ and $Z(3930)$. Using this setup, we predicted a $J^{PC}=0^{++}$ bound state corresponding to the $X(3860)$ by solving the coupled-channel Schr\"odinger equation.
The analysis of the resulting wave functions revealed the internal structure of these states. The $X(3872)$ is dominated by its molecular component, while the $X(3860)$ and $Z(3930)$ are primarily composed of the $c\bar{c}$ core. In the formation of the bound states, the coupling between the $c\bar{c}$ core and the molecular component plays a crucial role. 

The present model can be extended to study other exotic hadrons. The analysis of resonant states such as the $X(3915)$ and $X(4010)$ will be an important direction for future work.

\section*{Acknowledgement}
This work is supported by the RCNP Collaboration Research Network program as the project number COREnet-056.

\appendix
\section{\label{apd:obe}One-boson-exchange potential for $1^{++}$ and $2^{++}$.}
In this appendix, we summarize the potential matices of the pseudoscalar and vector meson exchanges for $J^{PC}=1^{++}$ and $2^{++}$.

\subsection{In the case of $1^{++}$}
\begin{gather}
	V_\pi        = \frac{1}{3}{\left(\frac{g}{2f_\pi}\right)}^2
	\begin{pmatrix}
		C_\pi           & 2C_\pi          & -\sqrt{2}T_\pi  & -2\sqrt{2}T_\pi & -\sqrt{6}T_\pi  & -2\sqrt{6}T_\pi \\
		2C_\pi          & C_\pi           & -2\sqrt{2}T_\pi & -\sqrt{2}T_\pi  & -2\sqrt{6}T_\pi & -\sqrt{6}T_\pi  \\
		-\sqrt{2}T_\pi  & -2\sqrt{2}T_\pi & C_\pi+T_\pi     & 2C_\pi+2T_\pi   & -\sqrt{3}T_\pi  & -2\sqrt{3}T_\pi \\
		-2\sqrt{2}T_\pi & -\sqrt{2}T_\pi  & 2C_\pi+2T_\pi   & C_\pi+T_\pi     & -2\sqrt{3}T_\pi & -\sqrt{3}T_\pi  \\
		-\sqrt{6}T_\pi  & -2\sqrt{2}T_\pi & -\sqrt{3}T_\pi  & -2\sqrt{3}T_\pi & C_\pi-T_\pi     & 2 C_\pi-2T_\pi  \\
		-2\sqrt{6}T_\pi & -\sqrt{6}T_\pi  & -2\sqrt{3}T_\pi & -\sqrt{3}T_\pi  & 2C_\pi-2T_\pi   & C_\pi-T_\pi
	\end{pmatrix},                                                                            \\
	V_\eta       = \frac{1}{9}{\left(\frac{g}{2f_\pi}\right)}^2
	\begin{pmatrix}
		C_\eta          & 0               & -\sqrt{2}T_\eta & 0               & -\sqrt{6}T_\eta & 0               \\
		0               & C_\eta          & 0               & -\sqrt{2}T_\eta & 0               & -\sqrt{6}T_\eta \\
		-\sqrt{2}T_\eta & 0               & C_\eta+T_\eta   & 0               & -\sqrt{3}T_\eta & 0               \\
		0               & -\sqrt{2}T_\eta & 0               & C_\eta+T_\eta   & 0               & -\sqrt{3}T_\eta \\
		-\sqrt{6}T_\eta & 0               & -\sqrt{3}T_\eta & 0               & C_\eta-T_\eta   & 0               \\
		0               & -\sqrt{6}T_\eta & 0               & -\sqrt{3}T_\eta & 0               & C_\eta-T_\eta
	\end{pmatrix},                                                                               \\
	V_{\lambda}  = \frac{1}{3}{\left(\lambda g_v\right)}^2
	\scriptsize
	\begin{pmatrix}
		-2C_\rho-2C_\omega               & -4C_\rho                         & -\sqrt{2}T_\rho-\sqrt{2}T_\omega   & -2\sqrt{2}T_\rho                   & -\sqrt{6}T_\rho-\sqrt{6}T_\omega   & -2\sqrt{6}T_\rho                   \\
		-4C_\rho                         & -2C_\rho-2C_\omega               & -2\sqrt{2}T_\rho                   & -\sqrt{2}T_\rho-\sqrt{2}T_\omega   & -2\sqrt{6}T_\rho                   & -\sqrt{6}T_\rho-\sqrt{6}T_\omega   \\
		-\sqrt{2}T_\rho-\sqrt{2}T_\omega & -2\sqrt{2}T_\rho                 & -2C_\rho+T_\rho-2C_\omega+T_\omega & - 4C_\rho+2T_\rho                  & -\sqrt{3}T_\rho-\sqrt{3}T_\omega   & -2\sqrt{3}T_\rho                   \\
		-2\sqrt{2}T_\rho                 & -\sqrt{2}T_\rho-\sqrt{2}T_\omega & -4C_\rho+2T_\rho                   & -2C_\rho+T_\rho-2C_\omega+T_\omega & -2\sqrt{3}T_\rho                   & -\sqrt{3}T_\rho-\sqrt{3}T_\omega   \\
		-\sqrt{6}T_\rho-\sqrt{6}T_\omega & -2\sqrt{6}T_\rho                 & -\sqrt{3}T_\rho-\sqrt{3}T_\omega   & -2\sqrt{3}T_\rho                   & -2C_\rho-T_\rho-2C_\omega-T_\omega & -4C_\rho-2T_\rho                   \\
		-2\sqrt{6}T_\rho                 & -\sqrt{6}T_\rho-\sqrt{6}T_\omega & -2\sqrt{3}T_\rho                   & -\sqrt{3}T_\rho-\sqrt{3}T_\omega   & -4C_\rho-2T_\rho                   & -2C_\rho-T_\rho-2C_\omega-T_\omega
	\end{pmatrix}, \\
	V_{\beta}    = {\left(\frac{\beta g_v}{2}\right)}^2
	\scriptsize
	\begin{pmatrix}
		-C_\rho/m_\rho^2-C_\omega/m_\omega^2 & -2C_\rho/m_\rho^2                    & 0                                    & 0                                    & 0                                    & 0                                    \\
		-2C_\rho/m_\rho^2                    & -C_\rho/m_\rho^2-C_\omega/m_\omega^2 & 0                                    & 0                                    & 0                                    & 0                                    \\
		0                                    & 0                                    & -C_\rho/m_\rho^2-C_\omega/m_\omega^2 & -2C_\rho/m_\rho^2                    & 0                                    & 0                                    \\
		0                                    & 0                                    & -2C_\rho/m_\rho^2                    & -C_\rho/m_\rho^2-C_\omega/m_\omega^2 & 0                                    & 0                                    \\
		0                                    & 0                                    & 0                                    & 0                                    & -C_\rho/m_\rho^2-C_\omega/m_\omega^2 & -2C_\rho/m_\rho^2                    \\
		0                                    & 0                                    & 0                                    & 0                                    & -2C_\rho/m_\rho^2                    & -C_\rho/m_\rho^2-C_\omega/m_\omega^2
	\end{pmatrix}.
\end{gather}

\subsection{In the case of $2^{++}$}
\begin{gather}
	V_\pi        = \frac{1}{3}{\left(\frac{g}{2f_\pi}\right)}^2
	\begin{pmatrix}
		0               & -2\sqrt{6/5}T_K   & -2\sqrt{6/5}T_K   & 2\sqrt{3}C_K      & 2\sqrt{3}C_K      & 2\sqrt{12/7}T_K   & 2\sqrt{12/7}T_K   \\
		-2\sqrt{6/5}T_K & C_\pi             & 2C_\pi            & -\sqrt{2/5}T_\pi  & -2\sqrt{2/5}T_\pi & \sqrt{14/5}T_\pi  & 2\sqrt{14/5}T_\pi \\
		-2\sqrt{6/5}T_K & 2C_\pi            & C_\pi             & -2\sqrt{2/5}T_\pi & -\sqrt{2/5}T_\pi  & 2\sqrt{14/5}T_\pi & \sqrt{14/5}T_\pi  \\
		2\sqrt{3}C_K    & -\sqrt{2/5}T_\pi  & -2\sqrt{2/5}T_\pi & -2C_\pi           & -4C_\pi           & \sqrt{4/7}T_\pi   & 2\sqrt{4/7}T_\pi  \\
		2\sqrt{3}C_K    & -2\sqrt{2/5}T_\pi & -\sqrt{2/5}T_\pi  & -42C_\pi          & -2C_\pi           & 2\sqrt{4/7}T_\pi  & \sqrt{4/7}T_\pi   \\
		2\sqrt{12/7}T_K & \sqrt{14/5}T_\pi  & 2\sqrt{14/5}T_\pi & \sqrt{4/7}T_\pi   & 2\sqrt{4/7}T_\pi  & C_\pi+3/7T_\pi    & 2C_\pi+6/7T_\pi   \\
		2\sqrt{12/7}T_K & 2\sqrt{14/5}T_\pi & \sqrt{14/5}T_\pi  & 2\sqrt{4/7}T_\pi  & \sqrt{4/7}T_\pi   & 2C_\pi+6/7T_\pi   & C_\pi+3/7T_\pi
	\end{pmatrix},                                                                            
        \end{gather}
    \begin{gather}
	V_\eta       = \frac{1}{9}{\left(\frac{g}{2f_\pi}\right)}^2
	\begin{pmatrix}
		0 & 0                 & 0                 & 0                 & 0                 & 0                 & 0                 \\
		0 & C_\eta            & 0                 & -\sqrt{2/5}T_\eta & 0                 & \sqrt{14/5}T_\eta & 0                 \\
		0 & 0                 & C_\eta            & 0                 & -\sqrt{2/5}T_\eta & 0                 & \sqrt{14/5}T_\eta \\
		0 & -\sqrt{2/5}T_\eta & 0                 & -2C_\eta          & 0                 & \sqrt{4/7}T_\eta  & 0                 \\
		0 & 0                 & -\sqrt{2/5}T_\eta & 0                 & -2C_\eta          & 0                 & \sqrt{4/7}T_\eta  \\
		0 & \sqrt{14/5}T_\eta & 0                 & \sqrt{4/7}T_\eta  & 0                 & C_\eta+3/7T_\eta  & 0                 \\
		0 & 0                 & \sqrt{14/5}T_\eta & 0                 & \sqrt{4/7}T_\eta  & 0                 & C_\eta+3/7T_\eta
	\end{pmatrix},                                                                                        \\
	\hspace{-5em} V_{\lambda}  = \frac{1}{3}{\left(\lambda g_v\right)}^2
	\tiny
	\begin{pmatrix}
		0                      & 2\sqrt{6/5}T_{K^{*}}                   & 2\sqrt{6/5}T_{K^{*}}                   & 4\sqrt{3}C_{K^{*}}                   & 4\sqrt{3}C_{K^{*}}                   & -2\sqrt{12/7}T_{K^{*}}                  & -2\sqrt{12/7}T_{K^{*}}                  \\
		2\sqrt{6/5}T_{K^{*}}   & 2C_\rho+2C_\omega                      & 4C_\rho                                & \sqrt{2/5}T_\rho+\sqrt{2/5}T_\omega  & 2\sqrt{2/5}T_\rho                    & -\sqrt{14/5}T_\rho-\sqrt{14/5}T_\omega  & -2\sqrt{14/5}T_\rho                     \\
		2\sqrt{6/5}T_{K^{*}}   & 4C_\rho                                & 2C_\rho+2C_\omega                      & 2\sqrt{2/5}T_\rho                    & \sqrt{2/5}T_\rho+\sqrt{2/5}T_\omega  & -2\sqrt{14/5}T_\rho                     & -\sqrt{14/5}T_\rho-\sqrt{14/5}T_\omega  \\
		4\sqrt{3}C_{K^{*}}     & \sqrt{2/5}T_\rho+\sqrt{2/5}T_\omega    & 2\sqrt{2/5}T_\rho                      & -4C_\rho-4C_\omega                   & -8C_\rho                             & -\sqrt{4/7}T_\rho-\sqrt{4/7}T_\omega    & -2\sqrt{4/7}T_\rho                      \\
		4\sqrt{3}C_{K^{*}}     & 2\sqrt{2/5}T_\rho                      & \sqrt{2/5}T_\rho+\sqrt{2/5}T_\omega    & -8C_\rho                             & -4C_\rho-4C_\omega                   & -2\sqrt{4/7}T_\rho                      & -\sqrt{4/7}T_\rho-\sqrt{4/7}T_\omega    \\
		-2\sqrt{12/7}T_{K^{*}} & -\sqrt{14/5}T_\rho-\sqrt{14/5}T_\omega & -2\sqrt{14/5}T_\rho                    & -\sqrt{4/7}T_\rho-\sqrt{4/7}T_\omega & -2\sqrt{4/7}T_\rho                   & 2C_\rho-3/7T_\rho+2C_\omega-3/7T_\omega & 4C_\rho-6/7T_\rho                       \\
		-2\sqrt{12/7}T_{K^{*}} & -2\sqrt{14/5}T_\rho                    & -\sqrt{14/5}T_\rho-\sqrt{14/5}T_\omega & -2\sqrt{4/7}T_\rho                   & -\sqrt{4/7}T_\rho-\sqrt{4/7}T_\omega & 4C_\rho-6/7T_\rho                       & 2C_\rho-3/7T_\rho+2C_\omega-3/7T_\omega
	\end{pmatrix}, \\
	V_{\beta}    = {\left(\frac{\beta g_v}{2}\right)}^2
	\tiny
	\begin{pmatrix}
		2C_\phi/m_\phi^2 & 0                                    & 0                                    & 0                                    & 0                                    & 0                                    & 0                                    \\
		0                & -C_\rho/m_\rho^2-C_\omega/m_\omega^2 & -2C_\rho/m_\rho^2                    & 0                                    & 0                                    & 0                                    & 0                                    \\
		0                & -2C_\rho/m_\rho^2                    & -C_\rho/m_\rho^2-C_\omega/m_\omega^2 & 0                                    & 0                                    & 0                                    & 0                                    \\
		0                & 0                                    & 0                                    & -C_\rho/m_\rho^2-C_\omega/m_\omega^2 & -2C_\rho/m_\rho^2                    & 0                                    & 0                                    \\
		0                & 0                                    & 0                                    & -2C_\rho/m_\rho^2                    & -C_\rho/m_\rho^2-C_\omega/m_\omega^2 & 0                                    & 0                                    \\
		0                & 0                                    & 0                                    & 0                                    & 0                                    & -C_\rho/m_\rho^2-C_\omega/m_\omega^2 & -2C_\rho/m_\rho^2                    \\
		0                & 0                                    & 0                                    & 0                                    & 0                                    & -2C_\rho/m_\rho^2                    & -C_\rho/m_\rho^2-C_\omega/m_\omega^2
	\end{pmatrix}.
\end{gather}
\bibliography{main}
\end{document}